\documentclass[showpacs,aps,graphicx,twocolumn]{revtex4}
\usepackage{amssymb}
\usepackage{graphicx}

\begin{document}

\title{Efficient quantum key distribution over a collective noise channel\footnote{Published in
Phys. Rev. A 78, 022321 (2008)}}
\author{ Xi-Han Li,$^{1,2,3}$
 Fu-Guo Deng,$^{1,3,4}$\footnote{Email address: fgdeng@bnu.edu.cn} and Hong-Yu Zhou$^{1,2,3}$}
\address{$^1$ The Key Laboratory of Beam Technology and Material
Modification of Ministry of Education, Beijing Normal University,
Beijing 100875, China\\
$^2$ Institute of Low Energy Nuclear Physics, and Department of
Material Science and Engineering, Beijing Normal University,
Beijing 100875, China\\
$^3$ Beijing Radiation Center, Beijing 100875,  China\\
$^4$ Department of Physics, Applied Optics Beijing Area Major
Laboratory, Beijing Normal University, Beijing 100875, China}
\date{\today }

\begin{abstract}
We present two efficient quantum key distribution schemes over two
different collective-noise channels. The accepted hypothesis of
collective noise is that photons travel inside a time window small
compared to the variation of noise. Noiseless subspaces are made
up of two Bell states and the spatial degree of freedom is
introduced to form two nonorthogonal bases. Although these
protocols resort to entangled states for encoding the key bit, the
receiver is only required to perform single-particle product
measurements and there is no basis mismatch. Moreover, the
detection is passive as the receiver does not switch his
measurements between two conjugate measurement bases to get the
key.
\end{abstract}
 \maketitle

\section{introduction}
Quantum key distribution (QKD) has become one of the most
important branches of quantum information. The principle of
quantum mechanics was introduced into communication  to ensure its
security by Bennett and Brassard (BB84) in 1984 \cite{bb84}, which
started the vigorous development of quantum communication.
Different from  classic communication, the security of quantum
communication is based on the laws of physics rather than the
difficulty of computation. The eavesdropper Eve is so powerful
that her ability is only limited by the principles in quantum
mechanics. However, the noncloning theorem forbids Eve to
eavesdrop the quantum signals freely and fully as her action will
inevitably disturb the unknown states and leave a trace in the
outcomes obtained by the two legitimate users. By far, QKD has
attracted the most attention
\cite{b92,Ekert91,BBM92,RMP,longqkd,CORE,BidQKD,Hwang,ABC}.

Photons are popular entities for quantum communication since they
are fast, cheap, easy to control and interact weakly with
environment. QKD experiments through free air and optical fibers
have been demonstrated over the past 20 years \cite{23,67,125}. It
is found in the experimental results that the polarization of
photons is incident to be influenced by the thermal fluctuation,
vibration, and the imperfection of the fiber, which are generally
called noise. Both the inhomogeneity of atmosphere in a free space
and the birefringence in an optical fiber are obstacles to the
application of quantum communication with photon polarization
degrees of freedom. The noise not only changes the fidelity of
quantum states carrying the information, which will decrease the
successful probability of schemes consequently, but also gives the
eavesdropper a chance to disguise her disturbance with a better
fiber, which will directly impact the key point of quantum
communication, i.e., its security. The most obvious solution is to
continuously estimate the transformation caused by the noise and
compensate for it momentarily. This can be denoted as a feedback
control project. However, this method is difficult in practice and
it requires an interruption of transmission. If the fluctuation is
too fast, the method is invalid.

There are two valid  methods to solve this problem: one is
choosing another degree of freedom to encode the key bits, and the
other is first to theorize the noise and then find a way to remove
or decrease the noise effect. The typical solution of finding a
new degree of freedom is phase coding, which has been demonstrated
in optical fibers experimentally \cite{67}. Although most of the
apparatus in the experiment was polarization-dependent, while the
polarization of the photon would be influenced by the
birefringence effect, the Faraday orthoconjugation effect
\cite{faraday} was proposed to circumvent this problem. However,
the phase-based schemes require complex interferometers and high
precision timing. Moreover, some phase coding protocols are
two-way communications that are susceptive to Trojan horse attack
\cite{RMP,attack}. The second method is first constructing an
appropriate noise model, and then finding a resolvent accordingly,
such as quantum error correct code (QECC)\cite{ecc}, single-photon
error rejection \cite{er1,er2}, quantum error-rejection code with
two qubits \cite{qerc1,qerc2,qerc3}, and decoherence-free subspace
(DFS) \cite{dfs1,dfs2,dfs3}. The QECC encodes one logical bit into
several physical qubits according to the type of noise, and then
the users measure the stabilizer codes to detect errors and
correct them. For single-photon error rejection schemes and
protocols utilizing the idea of DFS, there is an important
precondition called as the collective noise assumption
\cite{collective}. That is, the photons travel inside a time
window that is shorter than the variation of noise. In other
words, if several qubits transmit through the noise channel
simultaneously or they are close to each other spatially, the
transformation of the noise on each of the qubits is identical.

The single-photon error rejection schemes transmit photons
faithfully without ancillary qubits through a collective noise
channel. The two parts of the photon split by a polarizing beam
splitter (PBS) are adjusted to have a time delay, and then they
suffer from the same noise consecutively. The effect of noise is
cancelled by selecting the final state arriving at a special time
slot. In other words, the state collapses into a subspace that is
not  impacted by the noise with a certain probability. Kalamidas
proposed a single-photon error rejection protocol in 2005
\cite{er1}, which is efficient and convenient except for the use
of Pockels cells (PC). Recently, we presented a single-photon
transmission scheme with linear optics against collective noise
\cite{er2}, in which only passive linear optical elements are
required. In a sense, the schemes using only single-photon states
to reject errors can be regarded as a kind of DFS scheme in which
the time degree of freedom is introduced to form the DFS with the
two polarization parts.

The DFS can be made up of several qubits which experience the same
noise and compensate the effect of noise to implement a
fault-tolerance communication. Walton et al proposed a QKD scheme
using the idea of DFS in 2003 \cite{dfs1}. In their scheme, the
logical qubit is encoded into two time-bin qubits to protect the
quantum system against a collective-dephasing noise. The Hilbert
space is extended by the time degree of freedom, so that the
receiver could perform his measurement with a fixed basis. Later,
Boileau \emph{et al.} presented a QKD protocol with a collective
random unitary error model by using the linear combinations of two
singlet states $\vert \psi^- \rangle$, which are invariant under
whatever rotations \cite{dfs2}. The spatial degree of freedom is
also introduced to distinguish the states. However, the receiver
has to discard half of the samples due to the inconclusive
results, similar to traditional QKD protocols such as BB84 QKD
protocol in which the two legitimate users abandon half of the
outcomes owing to wrong measurement bases. Recently, Wang proposed
a robust QKD using a subspace of two-qubit states \cite{dfs3}. The
states being transformed out of the subspace by the collective
rotation are rejected by a parity check and then the total error
rate in the QKD protocol decreases.

In this paper, we present two fault-tolerant quantum key
distribution schemes against collective noises. One is used
against a collective-dephasing noise and the other is used against
a collective-rotation noise. The DFS is spanned by two entangled
states, and the spatial and polarization degrees of freedom are
both introduced. The receiver uses passive detection, i.e., he is
not required to switch between conjugate measurements, to obtain
the related outcomes. The most important merit of these two
schemes is that there is no basis mismatch, which means there is
no abandonment of samples owing to wrong basis measurement in
these two schemes.

\section{quantum key distribution against a collective noise}

We select special Bell states according to the form of noise to
build blocks for constructing a decoherence-free subspace. The key
bits are encoded on the states and the relative order of the
photon pairs.

\subsection{QKD against a collective-dephasing noise}
A collective-dephasing noise can be described as
\begin{eqnarray}
U \vert 0 \rangle = \vert 0 \rangle, \;\;\;\;\;\;\;\;  U \vert 1
\rangle = e^{i\phi} \vert 1 \rangle,
\end{eqnarray}
where $\phi$ is the parameter of the noise and it fluctuates with
time. Generally, the logical qubit encoded into two physical qubit
product states in the following is immune to this
collective-dephasing noise as the two logical qubits acquire the
same phase factor $e^{i\phi}$,
\begin{eqnarray}
\vert 0 \rangle_L = \vert 01 \rangle, \;\;\;\;\;\;\;\;  \vert 1
\rangle_L= \vert 10 \rangle,
\end{eqnarray}
where the subscript $L$ represents the logical bit, and $\vert 0
\rangle$ and $\vert 1 \rangle$ represent the horizontal
polarization state and  the vertical one, respectively, which are
the two eigenstates of Pauli operator $\sigma_z$ ($Z$ basis). We
choose two superpositions of these two logical bits to form a DFS.
They are two antiparallel Bell states written as
\begin{eqnarray}
\vert \psi^- \rangle &=& \frac{1}{\sqrt{2}}(\vert 01 \rangle -
\vert
10 \rangle),\\
\vert \psi^+ \rangle &=& \frac{1}{\sqrt{2}}(\vert 01 \rangle +
\vert 10 \rangle).
\end{eqnarray}

Generally speaking, a secure QKD protocol needs at least two
nonorthogonal measuring bases. The eavesdropper cannot obtain the
information directly and will disturb the quantum state without
the knowledge of its basis information. However, the use of two
nonorthogonal bases results in the abandonment of half instances
measured by the receiver with wrong bases, or calls for the
technique of quantum storage, which is difficult at present. For
constructing an efficient QKD protocol, we pack two two-particle
entangled states as one group and introduce the spatial degree of
freedom to form the nonorthogonality. The spatial degree of
freedom means the relative orders of the four particles. Two
permutations are used to form the two spatial bases to prepare the
quantum states, shown in Fig 1. There are the neighboring basis
$\vert \Psi \rangle$ in which the two entangled particles are in
proximity and the crossing basis $\vert \Phi \rangle$ in which
particles of the two entangled states are ranged alternately.

\begin{figure}[!h]
\begin{center}
\includegraphics[width=3.6cm,angle=0]{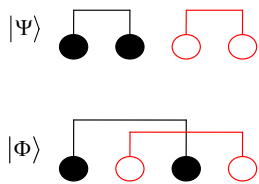} \label{f1}
\caption{Schematics of the two spatial bases. Lines represent the
entanglement between particles.}
\end{center}
\end{figure}

The four states with which we encode the  logical bits can be
written as
\begin{eqnarray}
\Psi^{dp}_0 &=& \vert \psi^+ \rangle_{12} \otimes \vert \psi^+
\rangle_{34}, \label{1}\\
\Psi^{dp}_1 &=& \vert \psi^- \rangle_{12} \otimes \vert \psi^-
\rangle_{34}, \label{2}\\
\Phi^{dp}_0 &=& \vert \psi^+ \rangle_{13} \otimes \vert \psi^+
\rangle_{24}, \label{3}\\
\Phi^{dp}_1 &=& \vert \psi^- \rangle_{13} \otimes \vert \psi^-
\rangle_{24}. \label{4}
\end{eqnarray}
Here the subscripts 0 and 1 on the left side represent the key
bits encoded, and subscripts $\{1,2,3,4\}$ on the right side
denote the sequence of these four particles on the line of
transmission. We can distinguish them by their time of arrival.
Under the assumption of collective noise, the interval between the
first and the fourth photons should be shorter than the
fluctuation time of the noise parameter $\phi$, which ensures that
these four photons suffer from the same noise.

These four states in terms of the X basis $\vert \pm x \rangle$
are shown below, where $\vert \pm x \rangle=
\frac{1}{\sqrt{2}}(\vert 0 \rangle \pm \vert 1\rangle)$ are the
two eigenstate of the Pauli operator $\sigma_x$. For the sake of
simplicity, we use $\vert \pm \rangle$ representing $\vert \pm x
\rangle$ in the following:
\begin{eqnarray}
\Psi^{dp}_0 = a+b,\\
\Psi^{dp}_1 = c-d,\\
\Phi^{dp}_0 = a+c,\\
\Phi^{dp}_1 = b-d,
\end{eqnarray}
where
\begin{eqnarray}
a &=& \frac{1}{2}(\vert ++++ \rangle + \vert ---- \rangle)_{1234},\\
b &=& \frac{1}{2}(\vert ++-- \rangle + \vert --++ \rangle)_{1234},\\
c &=& \frac{1}{2}(\vert +-+- \rangle + \vert -+-+ \rangle)_{1234},\\
d &=& \frac{1}{2}(\vert +--+ \rangle + \vert -++- \rangle)_{1234}.
\label{5}
\end{eqnarray}
It is not difficult to verify that $\langle \Psi^{dp}_0 \vert
\Psi^{dp}_1 \rangle =0, \langle \Phi^{dp}_0 \vert \Phi^{dp}_1
\rangle =0$ and $\langle \Phi^{dp}_i \vert \Psi^{dp}_i \rangle=
\frac{1}{2}$  $(i=0,1)$. For each basis $\Psi^{dp}$ and
$\Phi^{dp}$, the receiver can distinguish the two states in a
deterministic way with  four single-particle measurements.

Now, let us describe the QKD scheme in detail as follows:

(S1) The sender Alice chooses a random $(n+2\delta)$ bit string
$K$ and a random $(n+2\delta)$ bit string $B$.

(S2) Alice encodes each bit of the key string $K$ according to
$\{\vert \Psi^{dp}_0 \rangle,\vert \Psi^{dp}_1 \rangle\}$ if the
corresponding bit in the basis string $B$ is 0, or encodes into
$\{\vert \Phi^{dp}_0 \rangle,\vert \Phi^{dp}_1 \rangle\}$ if the
corresponding bit in  $B$ is 1.

(S3) Alice sends the $(n+2\delta)$ four-particle states to the
receiver Bob.

(S4) Bob performs the single-particle product measurements on each
quartet after the receipt subsequently. He selects randomly
$2\delta$ states from the sequence for eavesdropping check, where
$\delta$ states are measured with the $Z$ basis and the other
$\delta$ samples are measured in $X$ basis. The residual $n$ samples
used to share the secret key are measured in $X$ basis. Bob records
all of the measurement results.

(S5) Bob tells Alice the positions of groups chosen for
eavesdropping check and asks Alice for the initial states of these
samples. After receipt of these message, Bob checks the security
of the transmission by estimating the error rate. If the error
rate is acceptable, they continue to the next step. Otherwise,
they abort the protocol.

(S6) After they affirm  the security of the transmission, Alice
announces $B$, with which Bob can determine the key bits. These
results are taken as a raw key string. Error correction and
privacy amplification are required to obtain the final key.

Except for the samples used for the security analysis with the $Z$
basis measurements, all the other photons are measured with the
$X$ basis. That means the receiver is not required to switch
between conjugate measurement bases to get the message. And this
scheme is efficient as with Alice's information of the basis, all
the instances are used to generate the key, not just 1/4 of those
in the QKD scheme against a collective-dephasing noise shown in
Ref. \cite{dfs1}. Moreover, it does not require the receiver Bob
to measure his photons with joint two-photon measurements,
different from that in Ref. \cite{dfs1}.

\subsection{QKD against a collective-rotation noise}

Another common noise model called collective unitary rotation
noise satisfies
\begin{eqnarray}
U_r \vert 0 \rangle &=& cos \theta \vert 0 \rangle + sin \theta
\vert
1 \rangle\nonumber\\
U_r \vert 1 \rangle &=& -sin \theta \vert 0 \rangle + cos \theta
\vert 1 \rangle  .
\end{eqnarray}
The parameter $\theta$ depends on the noise in the quantum channel
and fluctuates with time. With such a type of collective errors,
we choose $\vert \psi^- \rangle$ and $\vert \phi^+ \rangle$ to
form the DFS, where
\begin{eqnarray}
\vert \phi^+ \rangle = \frac{1}{\sqrt{2}}(\vert 00 \rangle + \vert
11 \rangle).
\end{eqnarray}
They are invariant under this type of rotation. Similar to the
method used above, we pack two entangled states as one group. Four
combinations to code the key bits are
\begin{eqnarray}
\Psi^r_0 &=& \vert \phi^+ \rangle_{12} \otimes \vert \phi^+
\rangle_{34}=e+f, \label{r1}\\
\Psi^r_1 &=& \vert \psi^- \rangle_{12} \otimes \vert \psi^-
\rangle_{34}=g-h, \label{r2}\\
\Phi^r_0 &=& \vert \phi^+ \rangle_{13} \otimes \vert \phi^+
\rangle_{24}=e+g, \label{r3}\\
\Phi^r_1 &=& \vert \psi^- \rangle_{13} \otimes \vert \psi^-
\rangle_{24}=f-h, \label{r4}
\end{eqnarray}
where
\begin{eqnarray}
e &=& \frac{1}{2}(\vert 0000 \rangle + \vert 1111 \rangle)_{1234},\\
f &=& \frac{1}{2}(\vert 0011 \rangle + \vert 1100 \rangle)_{1234},\\
g &=& \frac{1}{2}(\vert 0101 \rangle + \vert 1010 \rangle)_{1234},\\
h &=& \frac{1}{2}(\vert 0110 \rangle + \vert 1001 \rangle)_{1234}.
\end{eqnarray}
It is easy to find that the superposition terms of $\{\Phi^r_0,
\Phi^r_1,\Psi^r_0,\Psi^r_1\}$ on the $Z$ basis is similar to those
of $\{\Phi^{dp}_0, \Phi^{dp}_1,\Psi^{dp}_0,\Psi^{dp}_1\}$ on the $X$
basis. In this QKD protocol, the sender Alice prepares a sequence of
quartet states randomly in $\Psi^r_0 (\Phi^r_0)$ or $\Psi^r_1
(\Phi^r_1)$ to denote the key bit 0 or 1. The choice of the basis
for each state is stochastic. Then she sends the sequence to Bob.
Bob chooses a sufficiently large subset of the multiplets as the
checking samples and measures them with the measuring bases $X$ and
$Z$ randomly. The other groups of states are measured with $Z$
basis. Bob records all of the measurement results and tells Alice
the positions of checking samples he selected. With Alice's
information of the original states of these checking samples, Bob
analyzes the error rate. If the error rate is reasonably low, they
determine the channel is secure. Otherwise, they abandon their
measurement results and repeat the communication from the beginning.
After ensuring the security of transmission, Alice tells Bob the
spatial bases she used to produce the quartets, with which Bob can
deduce the key sequence from his measurement outcomes.

\section{security analysis}

For one-way quantum communication, there are two main means of
eavesdropping. one is the intercept and resend attack
\cite{attack} and the other is that using an auxiliary particle to
interact with the particles carrying messages and measuring the
auxiliary photon to get some useful information. The interaction
can be two-particle unitary operation or controlled-not (CNOT)
gate \cite{ecc}. For simplicity, we denote the eavesdropping check
by measuring samples with the $X$($Z$) basis as the $X$($Z$) check
in the following.

Considering the first QKD scheme against a dephasing noise, as key
bits are encoded into four-particle entangled states, the direct
eavesdropping done by the eavesdropper Eve is the
intercept-measure-resend attack. There are two kinds of
measurements: the single-particle product measurements and the
Bell-basis measurements. For this scheme, the receiver Bob can
obtain the message only in the $X$ basis. If her measurement outcome
is a decomposition of $a$ $(d)$, Eve concludes that the key bit is
$0$ $(1)$ directly. For other results, Eve cannot get the key until
Alice publishes the information of spatial bases, $\Psi^{dp}$ or
$\Phi^{dp}$. If she resends a fake state(one of the four states,
$a,b,c,d$) to Bob according to her measurement results, the error
rate in $X$ check is $e_X=0$ and the error rate in $Z$ check is
$e_Z=$50\%. So the average error rate is $e_A=$25\%. There is
another way to resend the fake state. The eavesdropper Eve guesses
the original states according to her results and resends entangled
fake states. From Eqs. (\ref{1})-(\ref{4}), one can see that Eve
will get a wrong state with half opportunity, which will be
discovered with probability 25\% both in $X$ check and $Z$ check.
The average error rate of resending a guess state is 25\%. The
calculation of the error rate when Eve is measuring in the Bell
basis is a little complex because of the quantum entanglement
swapping phenomenon \cite{swap}. There is a half opportunity for Eve
to choose the correct spatial basis. For a wrong basis, Eve can
detect her mistake with 50\% due to the appearance of $\vert
\phi^{\pm} \rangle$ as results of entanglement swapping and then
change the basis to prepare a fake state. This method will cause a
25\% error rate in $X$ check and 12.5\% in $Z$ check, while the
average error rate is 19.25\%. It is important to point out that
with Bell-state measurements, Eve cannot get the key message until
the bases are published by Alice. The error rates of some
eavesdropping  are shown in Table I.

From Eqs. (\ref{1})-(\ref{5}), we find the parity of two particles
in the $X$ basis can reveal the key bit. For example, photons 3
and 4 are parallel in $\vert \Psi^{dp}_0 \rangle$ and antiparallel
in $\vert \Psi^{dp}_1 \rangle$. Photons 2 and 4 are parallel in
$\vert \Phi^{dp}_0 \rangle$ and antiparallel in $\vert \Phi^{dp}_1
\rangle$. The eavesdropper can utilize an auxiliary photon
prepared in $\vert + \rangle_5$ to get the key message by means of
two CNOT operators $C_{35}$ and $C_{45}$ along the $x$ direction,
where $C_{ij}$ means using particle $i$ as control bit and $j$ as
target bit. After the two CNOT gates, Eve measures photon 5 in the
$X$ basis. The outcome $\vert + \rangle (\vert - \rangle)$ means
the two photons 3 and 4 are parallel (antiparallel) and the key
bit is 0 (1). If Eve guesses the right spatial bases $\vert
\Psi^{dp} \rangle$ or $\vert \Phi^{dp} \rangle$, her action will
not be detected and she can get the key bit with the basis
information. Otherwise, this method will introduce a 50\% error
rate in $Z$ check.

\begin{table}[!h]
\footnotesize
\label{table1} \caption{The relation between the
error rate and the eavesdropping attack on the QKD protocol against
the dephasing noise. }

\begin{tabular}{ccccccccccccc}
\hline
 $dephasing~~~noise$ & & & & $e_X$ & & & & $e_Z$ & & &
&$e_A$\\\hline
 MB: X, resend: product state & &  & & 0 & & & & 50\% & & & &
 25\% \\
MB: X, resend: entangled state & &  & & 25\% & & & & 25\% & & & &
25\% \\
MB: Bell, resend: entangled state & &  & & 25\% & & & & 12.5\% & & &
& 19.25\% \\
  CNOT on auxiliary on X basis  & &  & & 0 & & & & 25\% & & & &
12.5\% \\
\hline
\end{tabular}
\end{table}

\begin{table}[!h]
\footnotesize \label{table2} \caption{The relation between the
error rate and the eavesdropping attack on the QKD protocol
against the rotation of polarization . }

\begin{tabular}{ccccccccccccc}
\hline
 $polarization~~~rotation$ & & & & $e_X$ & & & & $e_Z$ & & &
&$e_A$\\\hline
MB: X, resend: product state & &  & & 0 & & & & 50\% & & & &
 25\% \\
MB: X, resend: entangled state & &  & & 25\% & & & & 25\% & & & &
25\% \\
MB: Z, resend: product state & &  & & 50\% & & & & 0 & & & &
 25\% \\
MB: Z, resend: entangled state & &  & & 25\% & & & & 25\% & & & &
25\% \\
MB: Bell, resend: entangled state & &  & & 25\% & & & & 25\% & & &
&
25\% \\
CNOT on auxiliary on Z basis & &  & & 25\% & & & & 0 & & & &
12.5\% \\
\hline
\end{tabular}
\end{table}

For the second QKD scheme, the calculation of error rates for
different attacks is similar to the first one. A little difference
is caused by the fact that the states used in the second protocol
are symmetry in the two measuring bases (MBs) $X$ and $Z$. The
error rates are listed in Table II.
 From these two tables, we find that
the eavesdropper will introduce at least 12.5\% error inevitably
when she tries to wiretap. She will be detected by the two
legitimate users. Eve can  get half of the key bits both by $X$
measurement and $Z$ measurement with disturbing the unknown
quantum states, but Alice will announce the preparation bases
after the security check and the key bit will be used to encrypt
secret message after they confirm its security. So this QKD
protocol is secure in principle.

\section{discussion and summary}

Compared with the QKD scheme in Ref. \cite{dfs1}, our first QKD
scheme over a collective-dephasing noise channel has the advantage
of having a higher intrinsic efficiency \cite{cabello} as that in
the former is 1/4 and almost all the instances in our scheme can
be used to distill the private key. Moreover, this scheme requires
the receiver Bob only to perform single-photon measurements for
obtaining the outcomes used for distilling the private key and
does not require him to switch the choice of the measuring bases,
which will make the measurement simpler than that in Ref.
\cite{dfs1}. On the other hand, in order to encode the states
$\Psi^{dp}$ and $\Phi^{dp}$, the sender Alice needs to possess
some modulators for spatial modes, which will increase the
difficulty of the preparation of the logical qubit states. At
present, Alice can exploit the similar apparatus composed of
optical delays and switches in Ref. \cite{CORE} to adjust the
states $\Psi^{dp}$ and $\Phi^{dp}$. Also, Alice should prepare two
EPR pairs for each logical states. Although this task can be
accomplished at present by sending a pump pulse of ultraviolet
light  back and forth across a beta barium borate crystal
\cite{panjw,polarization}, it is not in a practical application
extensively.

Our second QKD scheme against a collective-rotation noise has a
higher intrinsic efficiency than that in the robust
polarization-based QKD scheme \cite{dfs2} as almost all the
instances in our scheme can be used to create the private key and
about 1/4 of the instances in the latter are useful (it can be
improved to be 1/2 if the proportion of the samples exploited by
the two parties to analyze the security of the quantum channel to
all the instances obtained is small). As the symmetry of the
encoding states shown in Eqs. (\ref{r1})-(\ref{r4}), we get a
scheme quite similar to a BB84 QKD protocol \cite{bb84} with
four-dimensional quantum systems. However, the scheme introduced
in Ref. \cite{dfs2} is more similar to the B92 QKD protocol
\cite{b92}. Through a channel with loss, our QKD scheme may be
more secure than that in Ref. \cite{dfs2}.

The idea of using only a few $X$ bases to check the eavesdropping
was proposed ten years ago \cite{asymmetry}. Its main aim is to
reduce the fraction of discarded data caused by wrong basis
measurement. There are some differences between it and our
schemes. First, in Ref. \cite{asymmetry} a predominant basis is
used to prepare and measure the states, but in ours a predominant
basis is only used to distill the key bits. That is, states
transmitting in the quantum channel are completely random in the
four states in our schemes,  so that a refined analysis of error
rate was required to prevent the eavesdropper from getting
information using the predominant basis in the former, while the
error rate analysis of ours is similar to the BB84 QKD protocol,
as is the security. Second, the samples for eavesdropping check
are chosen by Bob, which makes the process efficient as each one
selected is useful.

In summary, we propose two QKD schemes to share a sequence of key
with two different kinds of collective-noise channels. The
interference and two-way quantum communication is not required to
solve the problem of collective noises. Despite a little
difference in measuring basis and the states used to encode the
logical bits, the essence of these two schemes is quite similar.
That is, the legitimate user utilizes the special Bell states
which are invariant under the given noise model to protect the
system against the noise and introduces the spatial degree of
freedom to form two nonorthogonal bases with which the key rate is
increased compared to protocols abandoning half instances due to
wrong measuring bases. There are several remarkable advantages in
our two schemes.  First, the samples for security check are not
asked to be measured with an entangled basis. The samples are
chosen randomly by the receiver Bob which is easier, compared with
the QKD schemes with a decoy state, in which the sender Alice
inserts her decoy state into the massage sequence and tells Bob
the positions after the transmission. Second, it is unnecessary
for the two parties to discard samples. Except for eavesdropping
check, almost all of the states transmitted are used to share the
private key. Moreover, although the logical bits are encoded into
entangled states, the receiver only needs to perform
single-particle product measurements on his photons, not joint
two-particle Bell-state measurements. Except for the requirement
of two nonorthogonal bases measurement for eavesdropping check,
Bob need not switch his two conjugate bases to obtain the key.
Only passive detection is enough to get the message.

Schemes using several physical bits to present one logical bit are
fragile with photon loss and hence the communication distance is
restricted. This is a tradeoff between the transmission distance
and the degree of fault tolerance. More research and development
of technique are expected to solve this problem in the future.

\bigskip

\textbf{Note added} As the parity of the state of two EPR pairs
transmitted as a group can be detected with controlled-not
operations and an auxiliary particle, without disturbing the
quantum system, the two parties Alice and Bob can only exploit
each two EPR pairs to carry one bit of information securly over a
collective-noise channel. In the same way, the two parties in Ref.
\cite{CORE} should at least distill one bit of information about
the parity of each group of EPR pairs with privacy amplification;
otherwise, the eavesdropper Eve can obtain one bit of information
for each group of EPR pairs freely. With increasing of the number
of the EPR pairs in each group in Ref. \cite{CORE}, the
information leaked becomes less as each EPR pair carries two bits
of information in Ref. \cite{CORE}.

\section*{ACKNOWLEDGEMENTS}

This work is supported by the National Natural Science Foundation
of China under Grant No. 10604008, A Foundation for the Author of
National Excellent Doctoral Dissertation of China under Grant No.
200723, and  Beijing Natural Science Foundation under Grant No.
1082008.


\begin{thebibliography}{99}
\bibitem{bb84} C. H. Bennett and G. Brassard, in: Proceedings of IEEE
International Conference on Computers, Systems and Signal
Processing, Bangalore, India, IEEE, New York, 1984  p 175-179.

\bibitem{b92} C.H. Bennett, Phys. Rev. Lett. \textbf{68}, 3121 (1992).


\bibitem{Ekert91} A. K. Ekert, Phys. Rev. Lett.  \textbf{67},  661  (1991).

\bibitem{BBM92} C. H. Bennett, G. Brassard, and N. D. Mermin, Phys. Rev. Lett.
 \textbf{68},  557  (1992).

\bibitem{RMP} N. Gisin, G. Ribordy, W. Tittel, and H.
Zbinden, Rev. Mod. Phys.  \textbf{74}, 145  (2002).


\bibitem{longqkd} G. L. Long and X. S. Liu, Phys. Rev. A  \textbf{65},  032302
(2002).

\bibitem{CORE} F. G. Deng and G. L. Long, Phys. Rev. A
\textbf{68},  042315  (2003).

\bibitem{BidQKD} F. G. Deng and G. L. Long, Phys. Rev. A
 \textbf{70},  012311  (2004).

\bibitem{Hwang} W. Y. Hwang, Phys. Rev. Lett.  \textbf{91}, 057901  (2003).

\bibitem{ABC} H. K. Lo, H. F. Chau, and M. Ardehali, J. Cryptology  \textbf{18}, 133  (2005).

\bibitem{23} C. Kurtsiefer , P. Zarda, and M. Halder, Nature \textbf{419}, 450
(2002).

\bibitem{67} D. Stucki, N. Ginsin, O. Guinnard, and H. Zbinden, New.
J. Phys. \textbf{4}, 41 (2002).

\bibitem{125} X. F. Mo, B. Zhu, Z. F. Han, Y. Z. Gui, and G. C. Gan, Opt. Lett. \textbf{30}, 2632
(2005).

\bibitem{faraday} M. Martinelli, Opt. Commun. \textbf{72}, 341 (1989).

\bibitem{attack} F. G. Deng, X. H. Li, H. Y. Zhou, and Z. J.
Zhang, Phys. Rev. A \textbf{72}, 044302 (2005); X. H. Li, F. G.
Deng, and H. Y. Zhou, Phys. Rev. A \textbf{74}, 054302 (2006).

\bibitem{ecc}M. A. Nielsen and I. L. Chuang, Quantum Computation
and Quantum Information (Cambridge University Press,
Cambridge, England, 2000).

\bibitem{er1} D. Kalamidas, Phys. Lett. A \textbf{343}, 331, (2005).

\bibitem{er2} X. H. Li , F. G. Deng, and H. Y. Zhou, Appl. Phys. Lett.
\textbf{91}, 144101 (2007).

\bibitem{qerc1} X. B. Wang, Phys. Rev. Lett. \textbf{92}, 077902
(2004).

\bibitem{qerc2} X. B. Wang, Phys. Rev. A \textbf{69}, 022320
(2004).

\bibitem{qerc3} Y. A. Chen, A. N. Zhang, Z. Zhao, X. Q. Zhou, and J. W.
Pan, Phys. Rev. Lett. \textbf{96}, 220504 (2006).


\bibitem{dfs1} Z. D. Walton, A. F. Abouraddy, A. V. Sergienko, B. E. A. Saleh, and M. C. Teich,  Phys. Rev. Lett. \textbf{91},
087901 (2003).

\bibitem{dfs2} J. C. Boileau, D. Gottesman, R. Laflamme, D. Poulin, and R. W. Spekkens,  Phys. Rev. Let. \textbf{92},
017901 (2004).

\bibitem{dfs3} X. B. Wang, Phys. Rev. A \textbf{72}, 050304(R) (2005).

\bibitem{collective} P. Zanardi and M. Rasetti, Phys. Rev. Lett. \textbf{79}, 3306
(1997).

\bibitem{swap} M. Zukowski, A. Zeilinger,
M. A. Horne, and A. K. Ekert, Phys. Rev. Lett. \textbf{71}, 4287
(1993).

\bibitem{panjw} J. W. Pan, M. Daniell, S. Gasparoni, G. Weihs, and
A. Zeilinger, Phys. Rev. Lett. \textbf{86}, 4435 (2001).

\bibitem{polarization} C. Simon and J. W. Pan,  Phys. Rev. Lett. \textbf{89}, 257901 (2002).

\bibitem{asymmetry}  M. Ardehali, H. F. Chau, and Hoi-Kwong Lo, arXiv:
quant-ph/9803007


\bibitem{cabello} A. Cabello, Phys. Rev. Lett. \textbf{85}, 5635 (2000).

\end{thebibliography}
\end{document}